\documentclass[a4paper,fleqn]{cas-sc}

\usepackage[authoryear,longnamesfirst]{natbib}

\def\tsc#1{\csdef{#1}{\textsc{\lowercase{#1}}\xspace}}
\tsc{WGM}
\tsc{QE}
\tsc{EP}
\tsc{PMS}
\tsc{BEC}
\tsc{DE}
\newcommand{\GravC}{\mathcal{G}}

\begin{document}
\let\WriteBookmarks\relax
\def\floatpagepagefraction{1}
\def\textpagefraction{.001}
\shorttitle{Clustering and diffusion of TNOs}
\shortauthors{G.\ Pichierri and K.\ Batygin}
%\begin{frontmatter}

\title [mode = title]{Measuring the degree of clustering and diffusion of trans-Neptunian objects}

\author[1]{Gabriele Pichierri}[type=editor,
                        auid=000,bioid=1,
                        orcid=0000-0003-3622-8712]
\cormark[1]
\ead{gabe@caltech.edu, gabriele.pichierri@unimi.it}

\author[1]{Konstantin Batygin}[]

\affiliation[1]{organization={Division of Geological and Planetary Sciences, California Institute of Technology},
                addressline={1200 E.\ California Blvd}, 
                city={Pasadena},
                postcode={91125}, 
                state={CA},
                country={USA}}

\cortext[cor1]{Corresponding author}

\begin{abstract}
The outer solar system is populated by a broad aggregate of minor bodies, which occupy orbits whose dynamical character ranges from long-term stable to rapidly diffusive. We investigate the chaotic properties of known distant trans-Neptunian objects (TNOs) by numerically integrating TNO clones and statistically analyzing their orbital diffusion. Comparing the measured diffusion with an analytical criterion yields a dynamically motivated separation into classes of stable, metastable and unstable objects. We then measure the level of clustering of the longitudes of perihelia and of the orbital poles, as functions of orbital distance and of their stability properties. Distant (meta)stable objects appear increasingly clustered in perihelion around $\varpi \sim 50^\circ$ for increasing semi-major axis, while the orbits of unstable objects are well described by two, roughly equally-populated groups of ``clustered'' and ``anti-clustered'' objects, with means around $\sim 25^\circ$ and $\sim 205^\circ$ respectively.
We further find that, compared to the solar system's total angular momentum vector, the mean orbital poles of distant TNOs are significantly more misaligned for (meta)stable objects, while they remain roughly aligned for unstable objects. TNOs with intermediate orbital periods also appear to be misaligned with respect to the forced plane predicted by secular theory with the known planets. This gradation based on stability, if validated further by the upcoming VRO survey, necessitates a dynamical explanation.

\end{abstract}

\begin{keywords}
Kuiper belt \sep Trans-neptunian objects \sep Neptune
\end{keywords}

\maketitle

\section{Introduction} \label{sec:intro}
The trans-Neptunian objects (TNOs) are a population of minor bodies orbiting the Sun outside the orbit of Neptune and extending to thousands of astronomical units from the Sun. The existence of a belt of icy planetesimals at these orbital distances was predicted by \cite{1949MNRAS.109..600E} and \cite{1951PNAS...37....1K} as a consequence of formation processes of solids within the circumsolar nebula (see e.g.\ review by \citealt{2020tnss.book...25M}). After the singular discovery of Pluto in 1930, distant small bodies have been systematically observed starting from 1992's discovery of the Kuiper belt object $1992\text{QB}_1$ \citep{1993Natur.362..730J}, the first ``scattered disk'' object in 1997 \citep{1997Natur.387..573L} and Sedna's discovery in 2003 \citep{2004ApJ...617..645B}, just to name a few.
The fact that they are found on non-circular and inclined orbits, which is not the expected outcome of planetesimal formation, points to rich dynamical histories that sculpted the population over the lifetime of the Solar System. Indeed, unlike the cold classicals -- objects that reside on low-inclination orbits between (but outside) the 2:3 and 1:2 exterior mean motion resonances with Neptune --, many of these bodies' orbits are not primordial, and they are subject to a diverse set of dynamical mechanisms that perturb them in different ways depending on their orbital parameters. These include perturbations from the giant planets (close encounters, planetary scattering, various resonances, ...)\ and, for orbits with large enough orbital periods, perturbations external to the Solar System, such as galactic tides and passing stars, both during the cluster phase and in the field. The combination of these mechanisms generates a rich network of dynamical pathways to transport a TNO's orbit across various regions of the orbital parameter space (see e.g.\ review by \citealt{2020CeMDA.132...12S}).

While interesting in their own right, TNOs also represent a valuable testing ground for theories on the dynamical history of the known planets, of the solar cluster environment, and potentially hold the key to the discovery of additional unseen planets lurking in the outer region of the Solar System (e.g.,\ \citealt{2016AJ....151...22B,2017AJ....154...62V}). 
In this regard, a particularly intriguing property of the orbits of TNOs is the observed clustering of various orbital angles. \cite{2014Natur.507..471T} were the first to point out that the orbits with semi-major axis $a > 150$ AU (astronomical units) cluster in the argument of perihelion $\omega$. \cite{2016AJ....151...22B} subsequently pointed out that, among dynamically stable distant objects, the longitudes of perihelion $\varpi$ and of the node $\Omega$ also exhibit confinement. As an explanation for this pattern, they proposed that a yet-unseen distant planet with a mass a few times that of the Earth and on an eccentric, inclined orbit is responsible for creating and maintaining this clustering against the precession induced by the giant planets, which would erase such an alignment within a few hundred million years. 
The statistical significance of this clustering has been debated in the literature, with the works of \cite{2017AJ....154...50S}, \cite{2020PSJ.....1...28B} and \cite{2021PSJ.....2...59N} demonstrating that individual surveys (OSSOS, DES) cannot distinguish between underlying clustered and uniform populations. The global analysis of \cite{2019AJ....157...62B}, however, shows that the false-alarm probability associated with clustering in $\varpi$ and the angular momentum vectors is only $\simeq 0.2\%$.

Detection biases aside, orbital clustering appears to be regulated by dynamical stability. Objects whose orbits appear stable against perturbations from the giant planets (mainly Neptune) are typically clustered, while objects that appear unstable may reside on orbits with relatively random orientations (see review by \citealt{2019PhR...805....1B}). That said, a quantitative description of their stability and a more systematic analysis of these trends is lacking. Thus, regardless of the underlying mechanisms that produce these features, it is worth quantifying the degree of clustering of TNOs and how it relates to parameters such as orbital distance and the chaoticity of the objects.

In this paper, we propose such a scheme, and carefully measure the degree of stability of known distant TNOs by comparing their orbital diffusion (evaluated through gravitational integrations including the known giant planets) with a diffusion threshold that is set on theoretical grounds (Section \ref{sec:Da}). This leads to a dynamically motivated separation into non-diffusive, mildly diffusive and highly diffusive orbital states, which we denote as stable, metastable and unstable TNOs. We then statistically analyze the level of clustering of distant TNOs as a function of orbital distance, separating them by their stability properties (Sections \ref{sec:varpis} and \ref{sec:nodes} for clustering of perihelion longitudes and of orbital poles, respectively). The goal of this simple analysis is to place the empirical observation that stable objects tend to cluster more than unstable ones inside a quantitative and dynamically-motivated framework. Finally, we conclude in Section \ref{sec:conclusions}.
Although the dataset is arguably still limited at this point in time, the Vera C.\ Rubin Observatory (VRO) is expected to come online later in 2025, thus substantially increasing the number of TNOs and providing an independent view of the outer regions of the solar system with considerably better controlled and quantifiable biases. Our analysis is intended to set the groundwork to make sense of the upcoming wealth of data on the dynamical states of distant minor bodies in the solar system.

\section{Measuring diffusion in the TNO population} \label{sec:Da}

Our dataset of distant trans-Neptunian objects is based on a query of the small body database\footnote{
\url{https://ssd.jpl.nasa.gov/sb/}.
}. We integrate 25 clones of observed TNOs with $a>150$ AU, or TNOs with $a+\sigma_a > 150$ AU, except those with $\sigma_a > 100$ AU, so that highly undetermined orbits are excluded (we remark that the current census presents a biased view of the trans-Neptunian solar system region; see our discussion on biases below). Each orbital element in the set ${q, e, i, \Omega, \omega, \ell}$ (where $q$ is the perihelion distance, $e$ the eccentricity, $i$ the inclination, $\Omega$ the longitude of the ascending node, $\omega$ the argument of perihelion, and $\ell$ the mean anomaly) is selected from a normally distributed random variable to account for the uncertainty in the object's orbit; the semi-major axis $a=q/(1-e)$. Our simulations include all the giant planets, but neglect external perturbations such as galactic tide and passing stars, and are performed for up to 4 Gyrs, or until the particle is ejected from the system. To save on computational costs, we evolve multiple TNOs in the same simulation as non-interacting particles.

Instead of labeling each TNO's stability by visual inspection as in \cite{2019PhR...805....1B}, we take a more dynamically motivated approach. As a general rule, large-scale chaotic diffusion in phase space can be driven by overlap of isolated resonances \citep{1979PhR....52..263C}.
When the widths of isolated resonant islands are small enough that the latter are well separated in phase space, no large-scale diffusion is possible, and orbits can be described by appropriate action-angle variables where the actions (the momenta) remain constant (with thin, localized stochastic layers limited to small bands around the separatrices). When isolated resonances partially overlap, the momentum conjugate to the resonant angle can experience stochastic jumps within the region of overlap and some diffusion is possibles. Finally, when isolated resonances are fully overlapping (i.e.,\ the half-width of isolated resonances is approximatively equal to or larger than the distance between two adjacent resonances), strong diffusion over large portions of phase-space occurs in the direction perpendicular to the network of resonances.
In the case of scattered disk objects (SDOs) with $q > a_\mathrm{N}$, $a\gg a_\mathrm{N}$ (where quantities with subscript N refer to Neptune, and $a_\mathrm{N} = 30.1$ AU) and in the limit of coplanarity with Neptune, \cite{2021ApJ...920..148B} find that the relevant resonances that can overlap and cause scattering dynamics are the $2$:$k$ mean-motion resonances with Neptune. As the action conjugated to the resonant angle $\varphi_\mathrm{res} = k\ell - 2\ell_\mathrm{N} + 2\omega$ is proportional to $\sqrt{a}$, the semi-major axis of the small body can experience stochastic diffusion. The above classification in non-overlapping, partially overlapping and fully overlapping isolated resonances naturally produces three classes of non-diffusive (stable), mildly-diffusive (metastable) and strongly diffusive (unstable) orbits.
\cite{2021ApJ...920..148B} found that, within the stochastic layer of Neptune-driven diffusion, the Lyapunov time of scattered disk objects approaches the orbital period, while the semi-major axis diffusion coefficient is
\begin{equation}
    \mathcal{D}_a \sim \frac{8}{5\pi}\frac{m_\mathrm{N}}{M_\odot} \sqrt{\GravC M_\odot a_\mathrm{N}} \exp{[-(q/a_\mathrm{N})^2/2]}.
\end{equation}
They indeed show that this quantity measures the characteristic semi-major axis kick experienced by an SDO over a single orbital period (cfr.\ their equation (20) and subsequent discussion). From this, for a typical $q \gtrapprox 30$ AU, we choose a $\mathcal D_a$ of order $\mathcal D_{a,\mathrm{crit}} = 10^{-3}~\mathrm{AU}^2 \mathrm{yr}^{-1}$ as indicative of a diffusive evolution. Since  $\mathcal D_a$ is related to the change in semi-major axis by
\begin{equation}\label{eq:Da_def}
    \mathcal{D}_a \sim \frac{\delta a^2}{\tau},
\end{equation}
this yields a characteristic diffusion timescale $\tau \sim 10 - 100$ Myrs depending on the value of $q$. 

For each integrated TNO clone, we filter out the semi-major axis time evolution to eliminate short-term ($\sim$ Myrs) oscillations\footnote{
This is done using \texttt{scipy}'s function \texttt{uniform\_filter1d}. Tests using a cutoff of high-frequency oscillations in the Fourier spectrum of the signal gave similar results.
}; then, we measure the diffusion coefficient using equation (\ref{eq:Da_def}) along the simulation over the characteristic diffusion timescale $\tau$ chosen accordingly, and we take its maximum. For a given TNO, we then consider the average $\langle \mathcal{D}_a \rangle$ and standard deviation $\sigma_{\mathcal{D}_a}$ over all clones (after removing statistical outliers). We label a TNO as stable if $\langle \mathcal{D}_a \rangle < 0.5 \mathcal D_{a,\mathrm{crit}}$ and $\sigma_{\mathcal{D}_a} < \langle \mathcal{D}_a \rangle$; otherwise, if $\langle \mathcal{D}_a \rangle < 5 \mathcal D_{a,\mathrm{crit}}$ we label the object as metastable. In all other cases, we label the TNO as unstable. While the specific numerical choices are arbitrary, this scheme is intended to recognize orbits that are consistently non-diffusive as stable, marginally diffusive ones (or somewhat heterogeneously so) as metastable, and diffusive ones as unstable.

\begin{figure}[t!]
\includegraphics[width = 1. \textwidth]{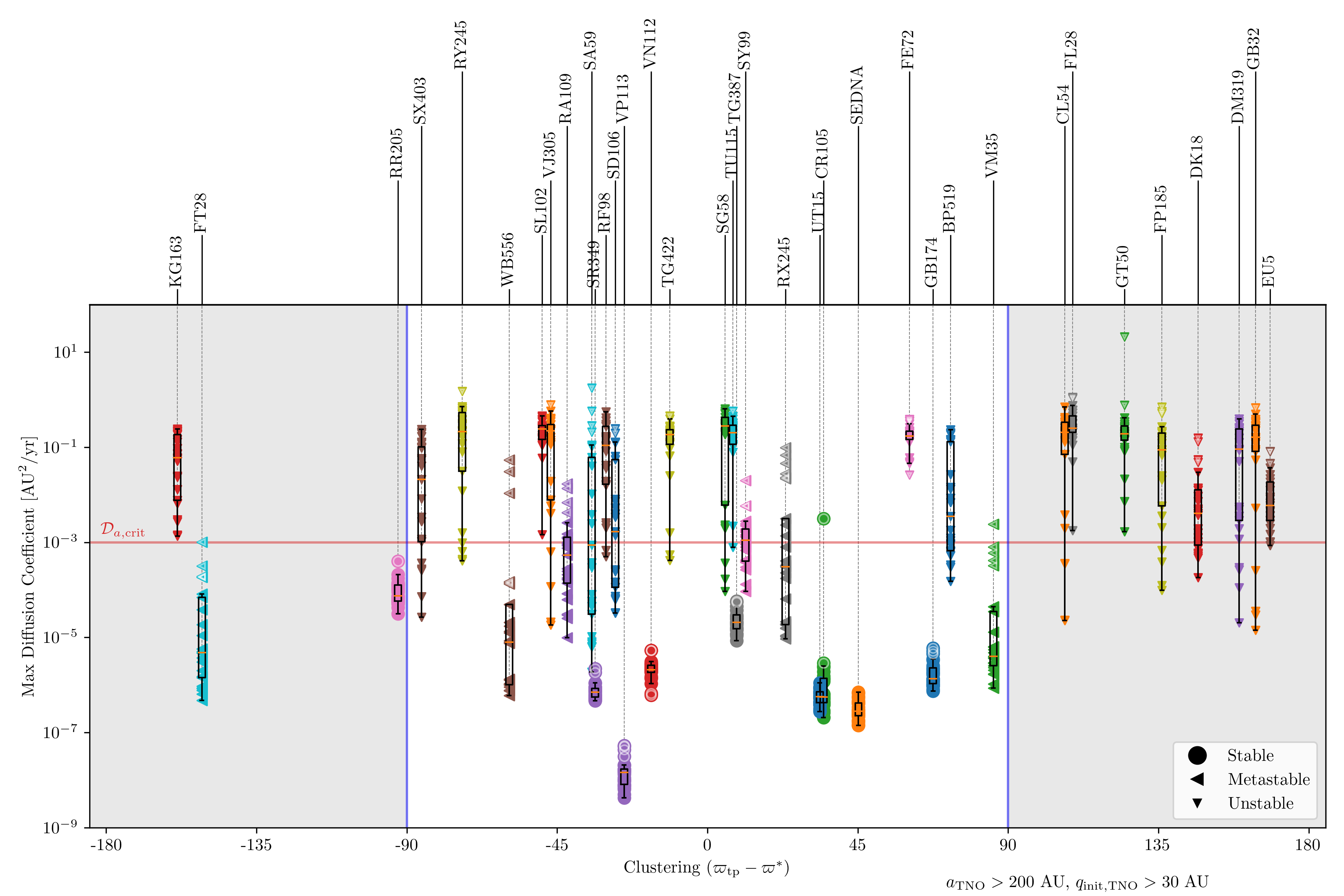}
\caption{Measure of orbital diffusion for a subset of the TNO sample ($a > 200$ AU, $q > 30$ AU) as a function of longitude of perihelion $\varpi$. Each TNO (see top labels) is represented by 25 clones, whose diffusion coefficients $\mathcal{D}_a$ are measured (vertical axis). Depending on the distribution of measured $\mathcal{D}_a$ with respect to a reference $\mathcal{D}_{a,\mathrm{crit}}$ (red horizontal line), the objects are labeled as stable, metastable and unstable (see bottom-right legend): in essence, TNOs whose $\mathcal{D}_a$ values are consistently above (below) the critical red line are labeled as unstable (stable), and they are labeled as metastable when they are in between the two regimes or with significant spread; see main text for details. The longitudes of perihelion are measured with respect to a reference value $\varpi^* = 50^\circ$ (Fig.\ \ref{fig:varpi_clustering_statistics}). Stable objects appear clustered around $\varpi^*$ within a region labeled with vertical blue lines, while unstable ones may reside outside this region (labeled as grey areas to the left and right of the plot).
\label{fig:varpi_clustering-vs-Da}}
\end{figure}

Figure \ref{fig:varpi_clustering-vs-Da} shows the measured $\mathcal{D}_a$ for all TNOs with $a > 200$ AU and $q>30$ AU as a function of their longitude of perihelion $\varpi = \omega + \Omega$ with respect to a $\varpi^* = 50^\circ$, a reference value chosen as motivated below. TNOs are labeled as stable, metastable or unstable according to the bottom-right legend. We also include a box plot to represent the statistical properties of each set of simulated TNOs; outliers are additionally labeled with white semi-transparent markers. A horizontal red line indicates the threshold value $\mathcal{D}_{a,\mathrm{crit}}$. We note that virtually all stable TNOs cluster within $\pm 90^\circ$ of $\varpi^*$. With the exception of $\mathrm{RR}_{205}$, which we flag as stable (although we note that it is the most diffusive among the stable TNOs here), all ``anti-clustered'' TNOs (meaning they fall more than $90^\circ$ away from $\varpi^*$) are unstable (with $\mathrm{FT}_{28}$ being metastable). We remark that these ``blocks'' of $180^\circ$ are arbitrary, but this represent the simplest (effectively binary -- ``clustered'' vs.\ ``anti-clustered'') counting measure. It is conceivable that, as the number of known objects grows, more subtle patterns will emerge and more sophisticated techniques and nomenclatures will become progressively warranted. As we still reside in the relatively low-number regime, however, we adopt the simplest approach.
We also note that, while other classifications of TNOs are in principle possible, we choose to make use of the degree of semi-major axis diffusion since this is ultimately driving their orbital (in)stability \citep{2021ApJ...920..148B}.

We show in the Appendix \ref{apx:A} how the TNOs project onto various orbital planes as a function of their stability properties assigned by our scheme. There, we find that (meta)stable and unstable objects are observed to coexist on the $a$ vs.\ $i$, $a$ vs.\ $\omega$ and $a$ vs.\ $\Omega$ planes. In terms of $a$ vs.\ $q$, stable objects are typically found to have larger pericenter distances than unstable ones; this is expected due to Neptune scattering efficiency depending on $q$ \citep{2020CeMDA.132...12S,2021ApJ...920..148B,2024MNRAS.527.3054H}. The threshold value of $q$ above which TNOs are stable depends on $a$, and is found to scale linearly with $\log_{10}(a/(1~\mathrm{AU}))$, but this trend roughly holds only for objects with $a\gtrsim200$ AU (Fig.\ \ref{fig:loga-vs-q__a-gtr-200_q-gtr-30__stab-through-mean-Da}).

We should also stress that the classification in terms of diffusion in phase space is based on simulations without an additional ninth planet \citep{2016AJ....151...22B,2019PhR...805....1B}. Adding a ``Planet Nine'' would in principle modify the stability properties of these objects by creating additional dynamical pathways to move in the space of orbital elements. 
However, we note that Planet-9-induced evolutions are secular, and happen on a much longer, Gyr-timescale. In fact, \cite{2019PhR...805....1B} find that the presence of a putative additional planet does not rapidly destabilize the objects' orbits. Indeed, only a small fraction of TNO clones that are labelled as stable are ejected, and the vast majority of objects that are stable without Planet Nine remain so with the addition of Planet Nine. 
In short, the presence of Planet Nine, at least with the orbital properties stated in \cite{2019PhR...805....1B}, is not expected to greatly modify the picture.
In any case, for future studies wishing to use the scheme presented here with the addition of a Planet Nine on a given orbit, we suggest that the stability of simulated TNOs (i.e.,\ their orbital diffusion) be analysed self-consistently.

\section{Distribution of perihelion longitudes}\label{sec:varpis}
We thus turn to a quantitative measurement of the degree clustering in TNOs and how it depends on their stability classification. For the longitude of perihelion $\varpi$, we use the von Mises distribution, an approximate analogue of the normal distribution which is suited for angular variables defined on the circle. The von Mises probability density function for an anglular variable $\theta$ (in radians) reads
\begin{equation}\label{eq:vM}
    f_\mathrm{vM}(\theta | \mu,\kappa) = \frac{\exp{[\kappa \cos(\theta-\mu)}]}{2\pi I_0(\kappa)},
\end{equation}
where $\mu$ is the mean and the parameter $\kappa$ is linked to the standard deviation of the distribution by $\sigma = \sqrt{1-I_1(\kappa)/I_0(\kappa)}$. Here $I_0$ and $I_1$ are the modified Bessel functions of first kind of order 0 and 1 respectively. For a given subset of TNOs, we use \texttt{scipy}'s \texttt{curve\_fit} to fit the functional form (\ref{eq:vM}) to the longitudes of perihelion, and extract $\mu$, $\kappa$ and thus $\sigma$, as well as the standard errors of the estimated parameters.

\begin{figure}[t!]
\includegraphics[width = 0.5\textwidth]{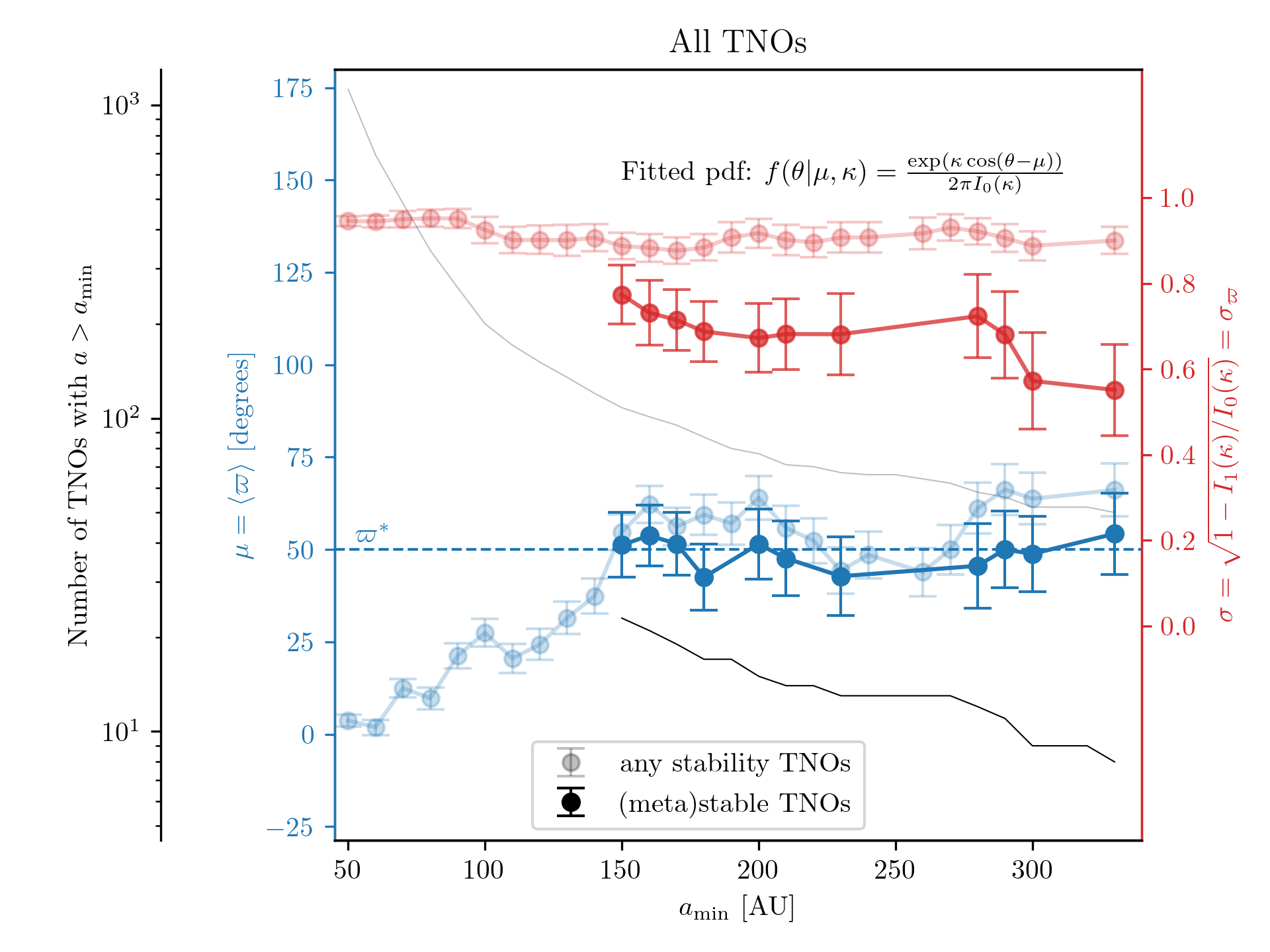}%
\includegraphics[width = 0.5\textwidth]{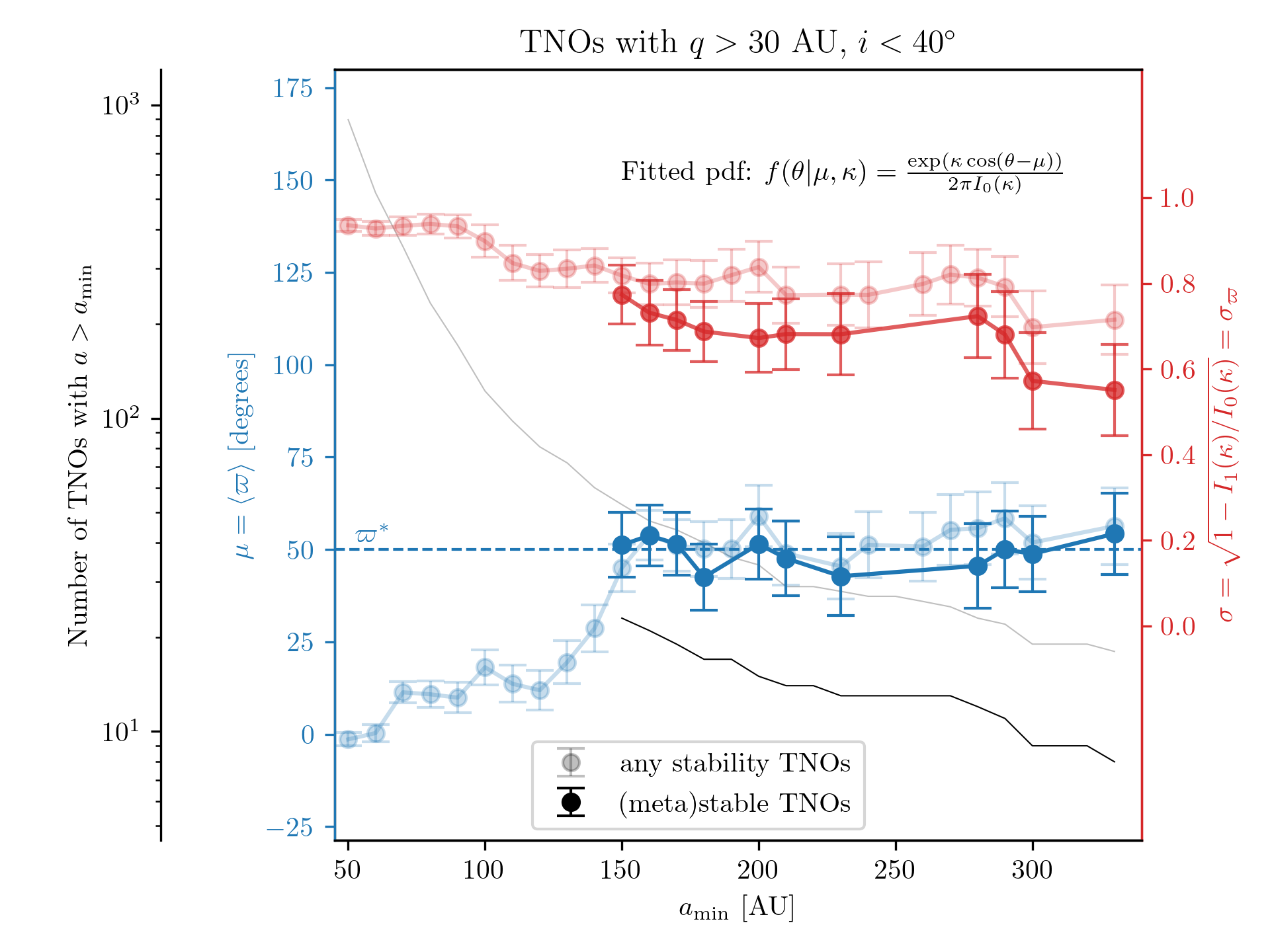}
\caption{Fitting the von Mises probability distribution function (Eq.\ (\ref{eq:vM})) to TNOs populations with $a > a_\mathrm{min}$ as a function of $a_\mathrm{min}$ (bottom axis).
In both panels, blue symbols represents the mean $\langle\varpi\rangle$ (see blue vertical axis on the left) and red symbols represent the standard deviation (red vertical axis on the right). Error bars represent the uncertainty in the parameter fit. Semi-transparent markers include all TNO's, while opaque markers only include (meta)stable ones (only for $a_\mathrm{min} > 150$ AU). The number of TNO's with $a>a_\mathrm{min}$ for both stability sets are indicated with black lines (see additional black vertical axis on the left). 
When the same subset of TNOs is obtained by imposing $a>a_\mathrm{min}$ for different subsequent values of $a_\mathrm{min}$, only one data point is shown for the $\mu$ and $\sigma$ statistics (this avoids repeating multiple times what is in fact the same data point).
Left panel: all TNOs are included. Right panel: only TNOs with $q>30$ AU and $i < 40^\circ$ are included. In general, for $a_\mathrm{min} \simeq 50$ AU objects show little clustering (high $\sigma$), while the mean of the perihelia approaches $\varpi^*\simeq 50^\circ$ (horizontal dashed line). For $a_\mathrm{min}>150$ AU, (meta)stable objects are significantly more clustered in perihelia around $\varpi^*$ than the full sample.
\label{fig:varpi_clustering_statistics}}
\end{figure}

In Figure \ref{fig:varpi_clustering_statistics}, for fixed values of $a_\mathrm{min}$ and for TNOs with semimajor axes $a > a_\mathrm{min}$, we plot in blue the mean $\langle\varpi\rangle$ and in red the standard deviation $\sigma_{\varpi}$ (note the different axes on the left and on the right with corresponding color schemes). Slightly transparent markers represent the fitted pdf parameters considering all TNOs with $a > a_\mathrm{min}$, while opaque markers indicate stable and metastable ones (which begin for $a_\mathrm{min} > 150$ due to our previous analysis). The left panel includes all TNOs, while the right panel is restricted to TNOs with $q>30$ AU and $i<40^\circ$. 
We group metastable objects together with the strictly stable ones as the number of the latter is too low and would lead to small-number statistics (note  the black curves, either semi-transparent or not, which track the number of TNOs used for the statistical analysis here, and the additional logarithmic axis on the left, also in black). We notice that for low $a_\mathrm{min}$, $\sigma_{\varpi} \gtrapprox 0.9$, indicating a broad distribution of perihelia with no clustering. At around $a_\mathrm{min} \simeq 125$ AU, the mean $\langle\varpi\rangle$ shifts to $\simeq 50^\circ$ which we thus set as our reference value $\varpi^*$. At the same time, the level of clustering rises as $\sigma_\varpi$ decreases. For $a_\mathrm{min} > 150$ AU we also note that the different levels of clustering becomes noticeable for the complete set of TNOs versus the set of (meta)stable ones, with the latter showing a higher level of clustering than the former: stable TNOs appear clustered around the mean $\varpi^*$ with a standard deviation $\sigma_\varpi \approx 0.5$, while the full set of TNOs shows a larger spread, with $\sigma_\varpi \gtrapprox 0.7$.

\subsection{Two distinct populations of diffusive TNOs}
\begin{figure}[t!]
\includegraphics[width = 1. \textwidth]{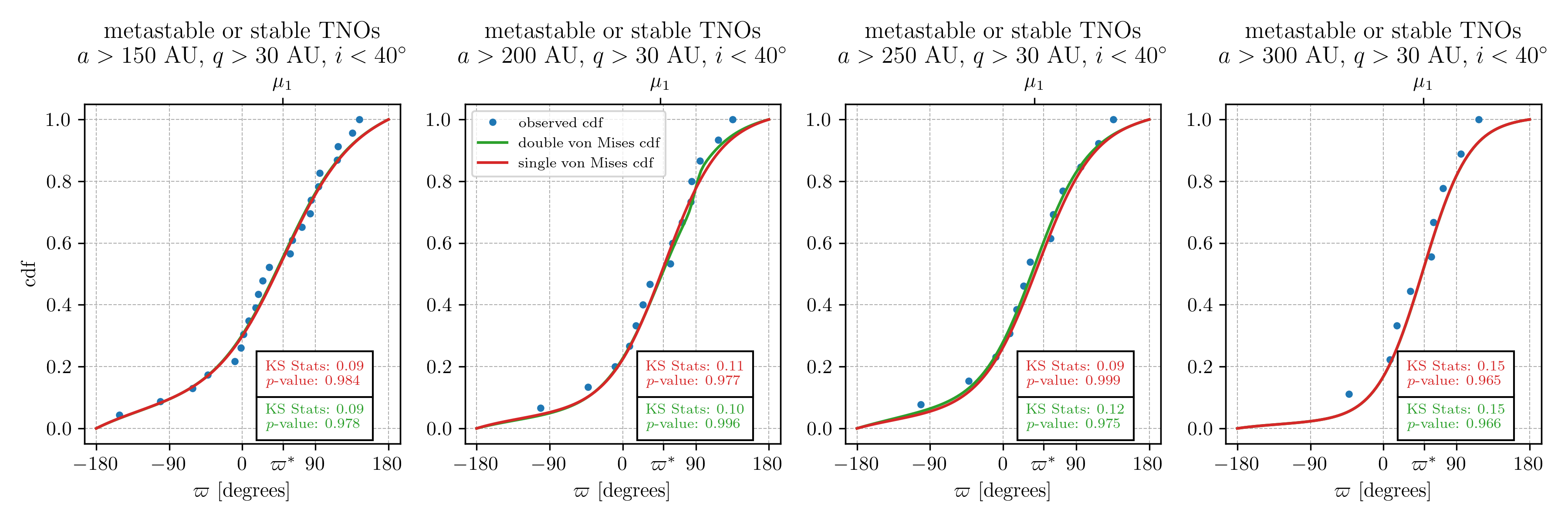}
\includegraphics[width = 1. \textwidth]{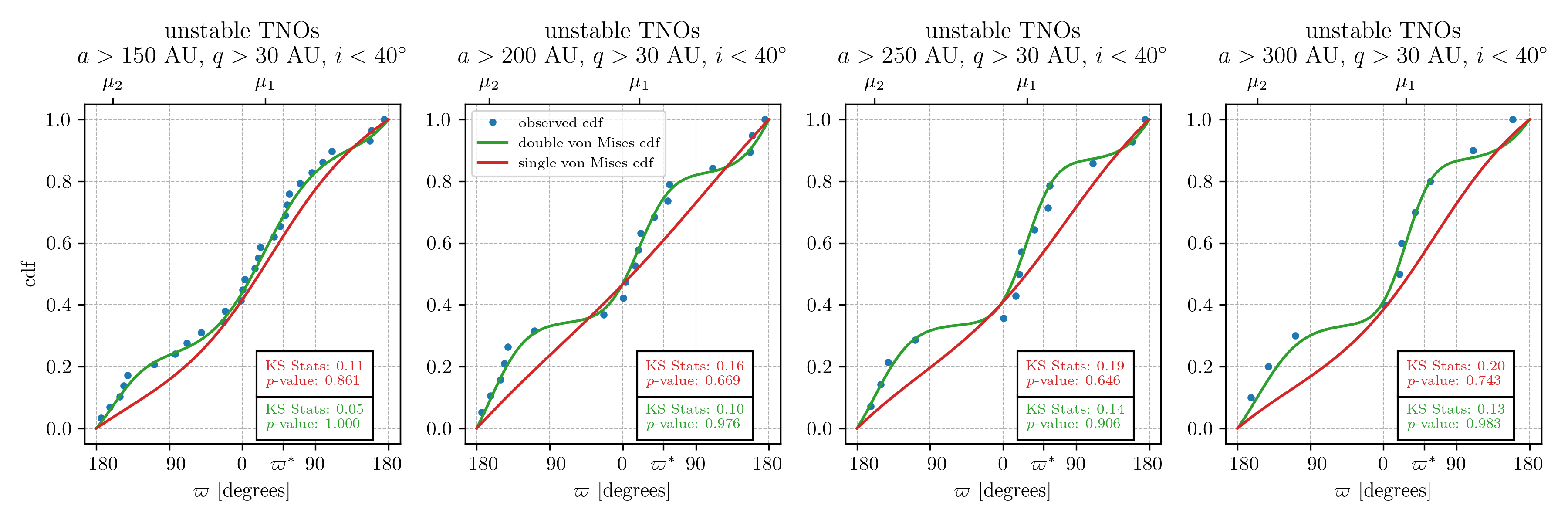}
\caption{Cumulative distribution functions for the longitude of perihelia of TNOs with $a > a_\mathrm{min}$ for 4 different $a_\mathrm{min}$ (different columns). TNOs are separated into (meta)stable and unstable (top and bottom row, respectively). The distribution of $\varpi$s of stable TNOs is consistent with a single population clustered around $\varpi^*$. Unstable TNOs are split into two equally-populated sets ($w\simeq 0.5$), one with $\mu_1 \simeq 25^\circ$ and the other one about $180^\circ$ away from it (which roughly fall within the ``clustered'' and ``anti-clustered'' regions in Fig.\ \ref{fig:varpi_clustering-vs-Da}, respectively).
\label{fig:varpi_clustering_pdfs__stats-Xstable_q-gtr-30_i-less-40}}
\end{figure}

\begin{figure}[ht!]
\includegraphics[width = 0.5\textwidth]{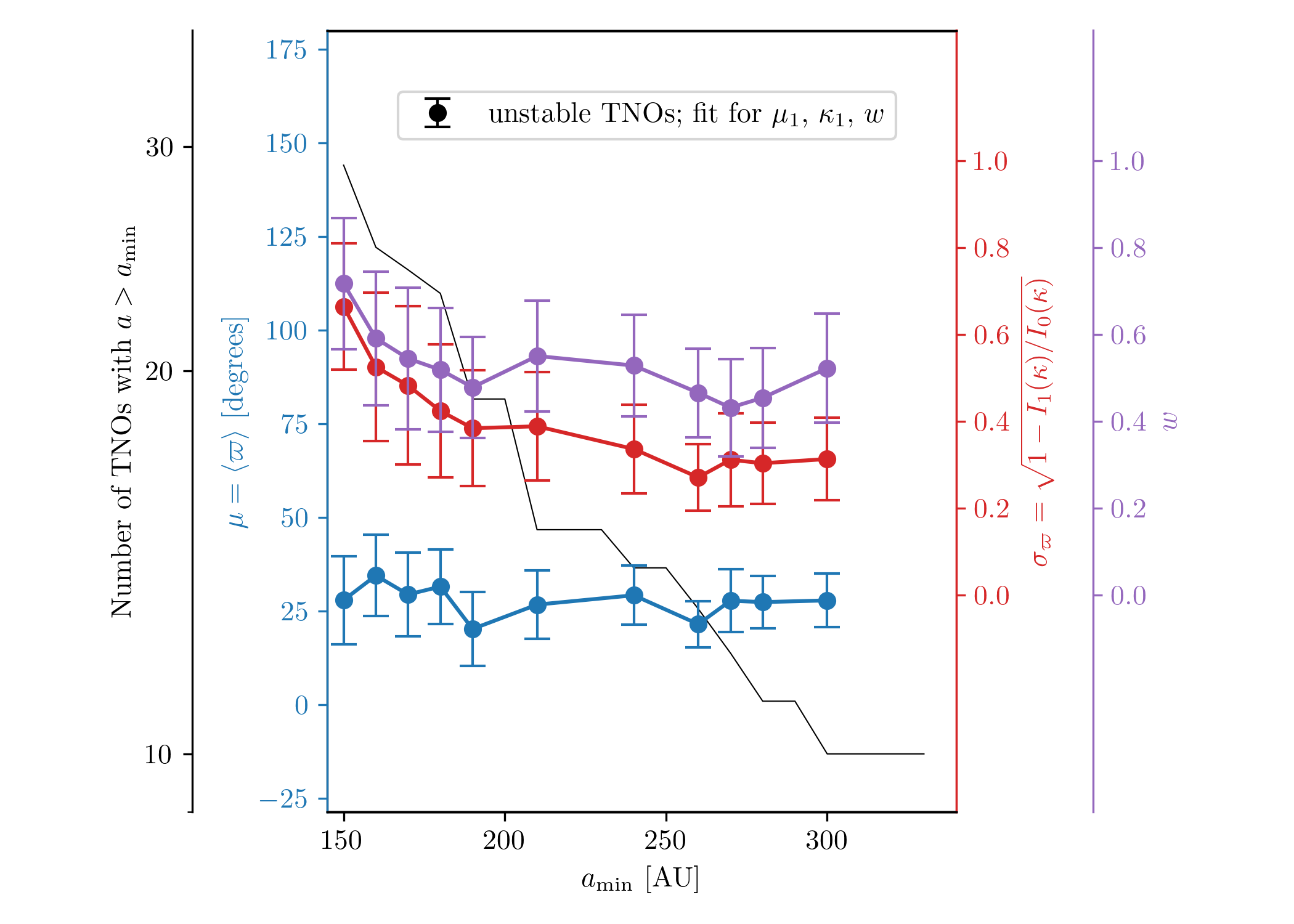}%
\includegraphics[width = 0.5\textwidth]{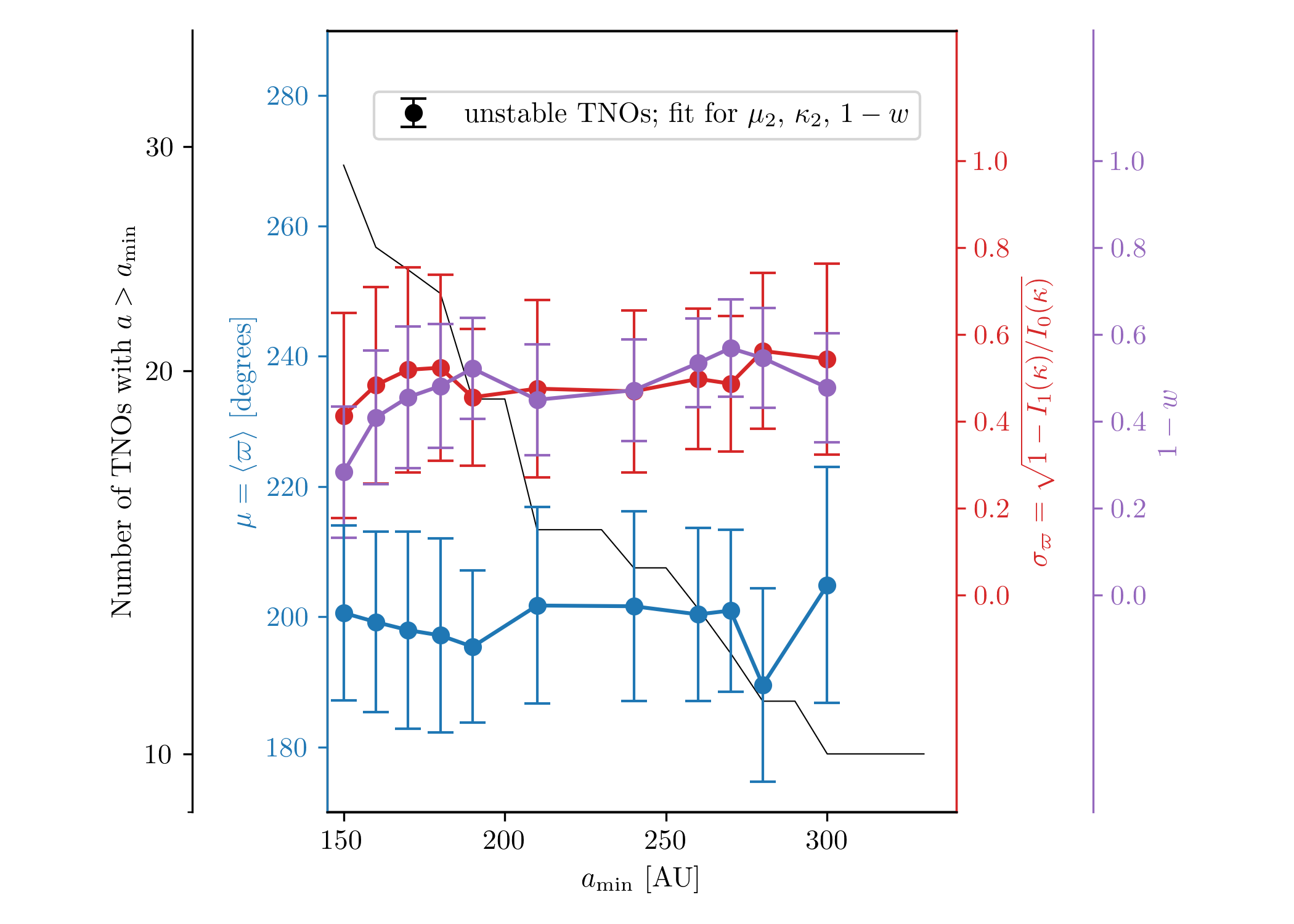}
\caption{Similar to Fig.\ \ref{fig:varpi_clustering_statistics}, but for a double von Mises distribution (Eq.\ (\ref{eq:2vM})); as in Fig.\ \ref{fig:varpi_clustering_statistics}, when we obtain the same subset of TNOs imposing $a>a_\mathrm{min}$ for different subsequent values of $a_\mathrm{min}$, only one data point is shown for the $\mu_{1,2}$, $\sigma_{1,2}$ and $w$ statistics. Left: Fitted parameters for the ``clustered'' population of unstable TNOs. Right: Fitted parameters for the ``anti-clustered'' population of unstable TNOs (notice the different vertical axis for $\langle\varpi\rangle$ in blue with respect to the left panel). In both panels, like in Fig.\ \ref{fig:varpi_clustering_statistics}, blue and red markers and their error bars represent the mean and standard deviations respectively (see left and right axes with corresponding colors), and the black lines indicates the number of objects with $a>a_\mathrm{min}$ (additional black axes on the left of each panel). Purple markers indicate the fit for $w$ (see additional purple axes on the right of each panel).
\label{fig:varpi_double-clustering_statistics__q-gtr-30_i-less-40}}
\end{figure}

Inspection of Figure \ref{fig:varpi_clustering-vs-Da} suggests that (meta)stable objects are typically clustered, while unstable ones may be either clustered or not. We thus test the hypothesis that unstable objects within the current census of TNOs follow a bimodal distribution of clustered and anti-clustered populations.
We fit a weighted linear combination of von Mises distributions:
\begin{equation}\label{eq:2vM}
    f_\mathrm{2vM}(\theta | \mu_1,\kappa_1,\mu_2,\kappa_2,w) = w f_\mathrm{vM}(\theta | \mu_1,\kappa_1) + (1-w)f_\mathrm{vM}(\theta | \mu_2,\kappa_2),
\end{equation}
where $w\in[0,1]$ is a weight parameter. We use the same method as above and fit the free parameters $\mu_1,\kappa_1,\mu_2,\kappa_2,w$ to the distributions of TNOs with $a>a_\mathrm{min}$, $q>30$ and $i<40^\circ$, separating the unstable objects from the rest (metastable and stable). For different values for $a_\mathrm{min}$, we find that the distribution of perihelia of the metastable and stable objects with $a>a_\mathrm{min}$ are consistently well reproduced by a single von Mises distribution ($w\approx 1$), with mean $\mu_1\simeq\varpi^*$ (Fig.\ \ref{fig:varpi_clustering_pdfs__stats-Xstable_q-gtr-30_i-less-40}, top panel). Instead, the distribution of perihelia of unstable TNOs with $a>a_\mathrm{min}$ is bimodal and very well fitted by a double von Mises distribution as in (\ref{eq:2vM}), with $w\approx 0.5$ (Fig.\ \ref{fig:varpi_clustering_pdfs__stats-Xstable_q-gtr-30_i-less-40}, bottom panel). Figure \ref{fig:varpi_double-clustering_statistics__q-gtr-30_i-less-40} is similar to Figure \ref{fig:varpi_clustering_statistics} but shows the separate fits for $\mu_1$, $\kappa_1$ and $\mu_2$, $\kappa_2$ (panels a and b, respectively) as well as the weighting parameter $w$. We find that for $a_\mathrm{min} > 200$ AU, $\mu_1\simeq 25^\circ$, $\mu_2 \simeq 205^\circ$, and $w$ is approximately 0.5, indicating two separate and equally populated subgroups. Interestingly, $\mu_1$ is more than $1\sigma_\varpi$ away from the mean $\varpi^*$ for the overall population (Fig.\ \ref{fig:varpi_clustering_statistics}); however, $\mu_1\simeq 25^\circ$ still falls within the clustered population in Figure \ref{fig:varpi_clustering-vs-Da}. Keeping in mind that the $\pm 90^\circ$ bounds are somewhat arbitrary in separating the full $[0^\circ,360^\circ]$ range into two halves, we may interpret the unstable set following the von Mises distribution associated to $\mu_1$, $\kappa_1$ as part of the clustered population. The other set, associated to $\mu_2$, $\kappa_2$, is centered roughly $180^\circ$ away from the cluster, at $\mu_2 \simeq 205^\circ$, and appears somewhat more spread out than the clustered set, with a dispersion about twice as large. 
Assuming that these dynamical signatures are due to the presence of an unseen planet, this is consistent with a secular evolution in the $e$ vs.\ $\Delta\varpi = \varpi-\varpi_\mathrm{pert}$ plane driven by an external perturber (e.g.,\ Fig.\ 10 of \citealt{2019PhR...805....1B}). Objects that follow the level curves of the secular Hamiltonian around the apsidally-aligned equilibrium reach high eccentricities and thus more strongly interact with Neptune when their longitude of perihelion is about $30^\circ$ away from exact apsidal alignment. Note that, together with the chaotic diffusion of semi-major axis and on top of any putative additional perturbations, the orbital elements of TNOs also undergo quasi-secular evolution.

\section{Distribution of orbital planes}\label{sec:nodes}
\begin{figure}[ht!]
\centering
\includegraphics[width = 0.75\textwidth]{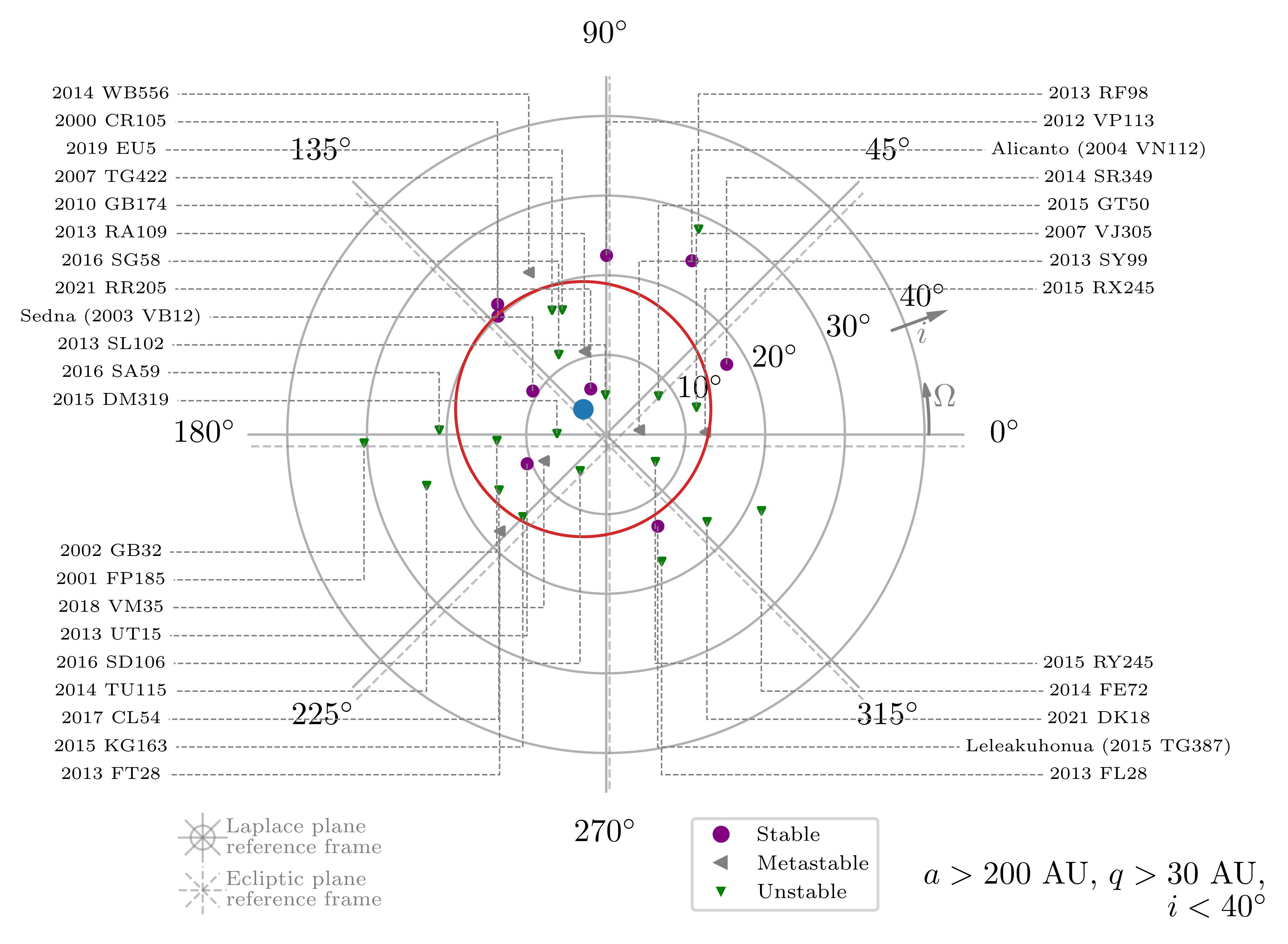}
\caption{
Distribution of orbital poles of distant TNOs ($a>200$ AU, $q>30$ AU, $i < 40^\circ$) with respect to the invariable plane and the ecliptic plane (see bottom left legend). Each TNO is labeled by its stability property in a similar fashion to Fig. \ref{fig:varpi_clustering-vs-Da}. The mean and standard deviation for this subset of TNOs are shown as a blue dot and a red circle.
\label{fig:ang_mom_polar__a-gtr-200_obs_with-EP}}
\end{figure}

\begin{figure}[ht!]
\includegraphics[width = 0.5\textwidth]{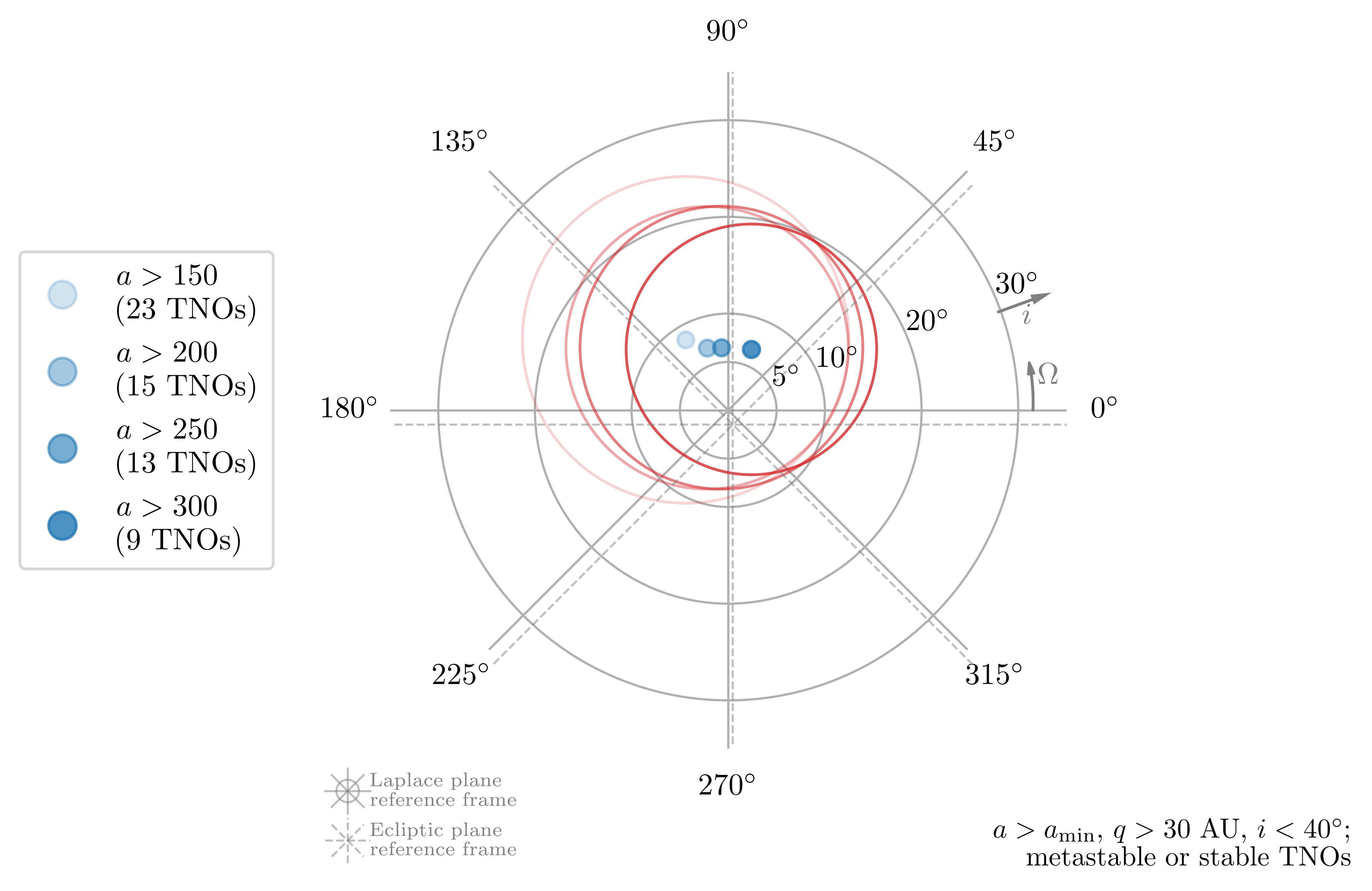}%
\includegraphics[width = 0.5\textwidth]{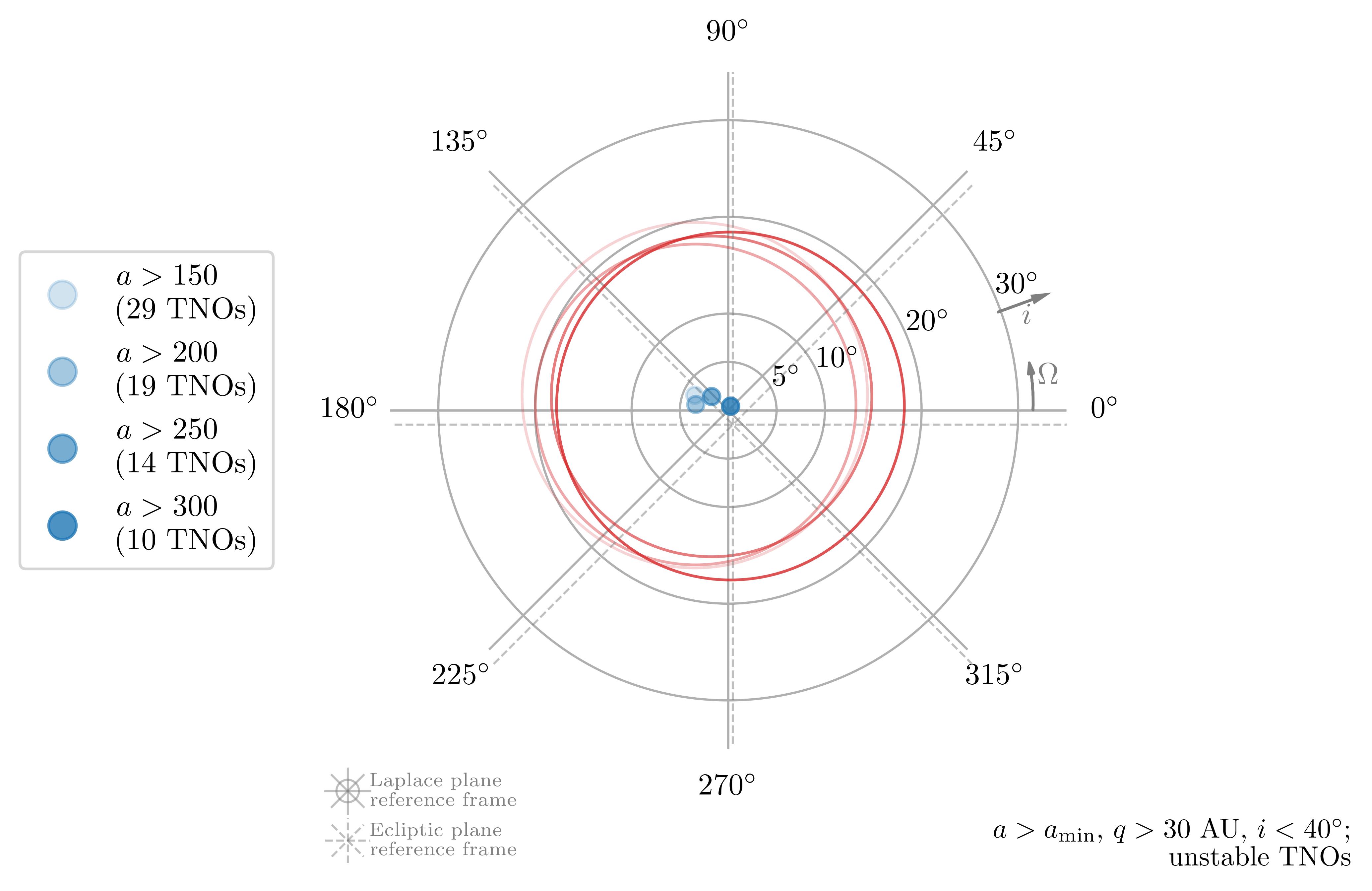}
\caption{
Warping of the mean orbital plane of distant TNOs with $a>a_\mathrm{min}$ for different values of $a_\mathrm{min}$ (see legend on the left). Blue dots and red circles represent the mean and standard deviation for the subset of TNOs considered, labeled graphically by different opacities.
Left: planes of (meta)stable TNOs. Right: planes of unstable TNOs.
\label{fig:ang_mom_polar__stats-Xstable_w_EP}}
\end{figure}

The orbital planes of distant TNOs also appear clustered away from both the ecliptic and the Laplace invariable plane (itself tilted with respect to the ecliptic by $i_\mathrm{ip}\simeq 1.58^\circ$ and $\Omega_\mathrm{ip} \simeq 107.58^\circ$; e.g.,\ \citealt{2012A&A...543A.133S}). Figure \ref{fig:ang_mom_polar__a-gtr-200_obs_with-EP} shows the orbital poles $(i \cos\Omega, i \sin\Omega)$ for TNOs with $a>250$ AU, $q>30$ AU and $i<40^\circ$, labeled by their stability (similar to Fig.\ \ref{fig:varpi_clustering-vs-Da}). Unlike the longitude of perihelia, we statistically describe the distribution of poles simply by their average (blue dot in Fig.\ \ref{fig:ang_mom_polar__a-gtr-200_obs_with-EP}) and the spread around the mean (red circle) on the $(i \cos\Omega, i \sin\Omega)$ plane.

Figure \ref{fig:ang_mom_polar__stats-Xstable_w_EP} shows the mean and standard deviation of the poles of (meta)stable (left panel) and unstable (right panel) TNOs with $a > a_\mathrm{min}$, with different values for $a_\mathrm{min}$. We again see that distant (meta)stable objects appear to be significantly more misaligned ($i \simeq 5^\circ - 10^\circ$) with respect to the invariable plane than unstable ones, which for large enough semi-major axes are instead perfectly aligned with it.

\begin{figure}[ht!]
\centering
\includegraphics[width = 0.5\textwidth]{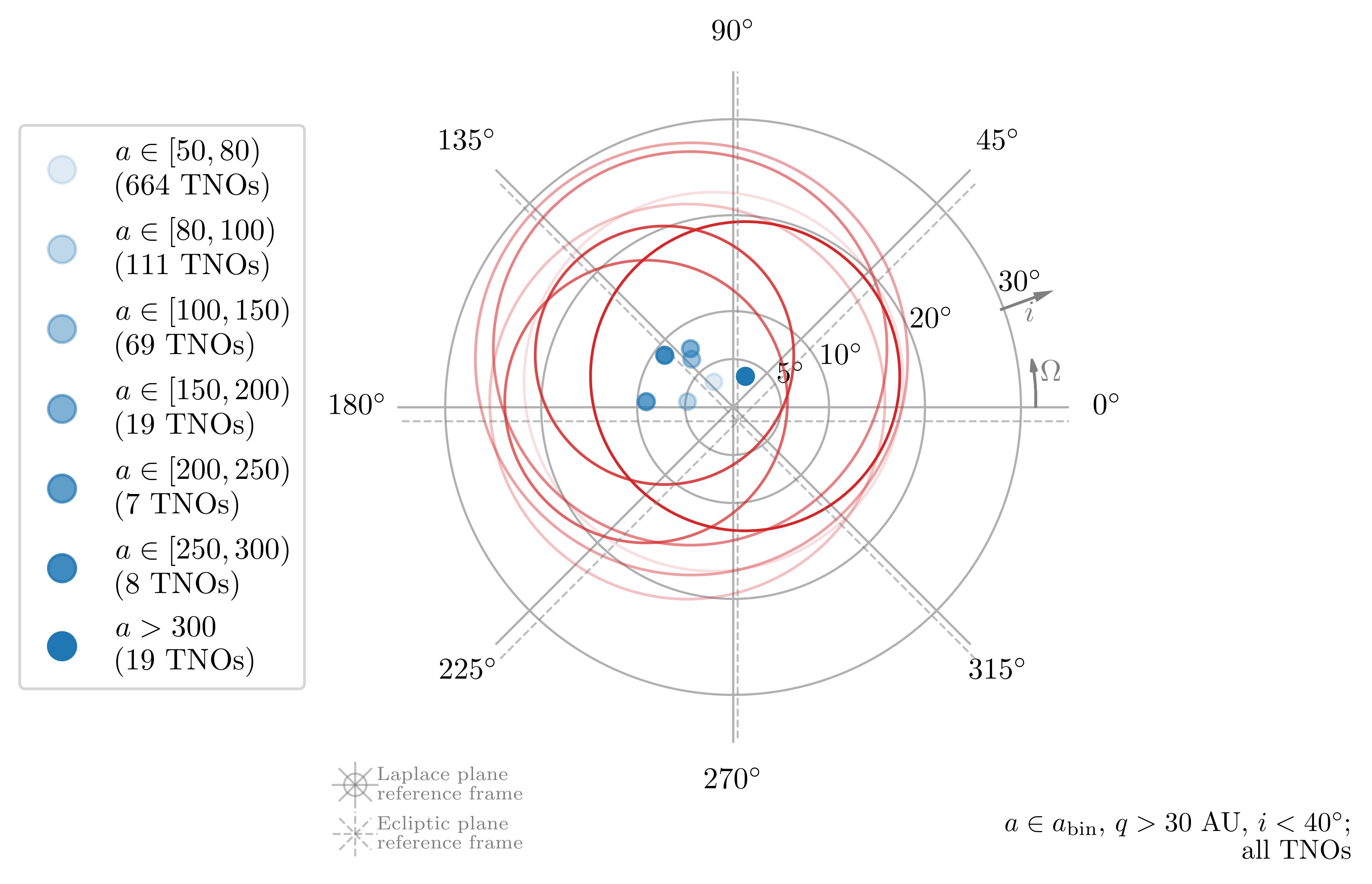}
\caption{Similar to Fig. \ref{fig:ang_mom_polar__stats-Xstable_w_EP}, but grouping TNOs in bins within a given range of semi-major axes. Taking the data at face value shows that even for semi-major axes between $\simeq 50$ and $\simeq 150$ AU the orbital planes are warped by a few degrees.
\label{fig:ang_mom_polar__binning_all_w_EP_from50AU}}
\end{figure}

Interestingly, when we consider objects that are closer in, the average plane appears to remain warped with respect to the invariable plane, which would represent the forced plane around which orbits  precess down to $a\simeq 50$ AU in a purely secular theory considering all the (known) planets in the Solar System. 
This feature was already observed in \cite{2017AJ....154...62V}, who argued that an unseen, relatively close-in, low-mass planet could explain the warping. However, \cite{2019AJ....158...49V} argued that OSSOS data rejects the hypothesis of a misalignment of the forced plane based on the known planet with the invariable plane for $a \simeq 50$ to 150 AU. Further out, \cite{2020PSJ.....1...28B} observe a $\sim2.2\sigma$ warp in $\Omega$ for $a>250$ AU in the DES sample, although they do not rule out the hypothesis that the underlying population is in fact isotropic based on the small-number statistics. We note however that $\Omega$ looses meaning for vanishing $i$, so that one ultimately needs to look at the two-dimensional quantity $(i \cos\Omega, i \sin\Omega)$, as in Figures \ref{fig:ang_mom_polar__a-gtr-200_obs_with-EP}, \ref{fig:ang_mom_polar__stats-Xstable_w_EP} and \ref{fig:ang_mom_polar__binning_all_w_EP_from50AU}.

If scattered-disk objects are constantly injected from a misaligned inner-Oort cloud, the observed warp could be explained by the original misalignment of the latter. The giant-planet-driven secular precession of the node, given by
\begin{equation}\label{eq:dotOmega}
\dot\Omega = -\frac{3}{4}\sqrt{\frac{\GravC M_\odot}{a^3}} \frac{\cos(i)}{(1-e^2)^2} \sum_{j=5}^8 \frac{m_j}{M_\odot}\left(\frac{a_j}{a}\right)^2
\end{equation}
(where $a_j$ and $m_j$ are the semi-major axes and masses of the giant planets),
would erase any clustering over a timescale of order of a few hundred Myrs. This would require the removal of injected TNOs over a shorter timescale to maintain a signature of the original alignment of the poles. However, the timescale for the precession of the longitude of the perihelia is only a factor $\cos(i)$ from (\ref{eq:dotOmega}), and the mean of $\varpi$ does not seem to remain highly clustered around $\varpi^*$ for $a \simeq 50$ to 150 AU (Fig.\ \ref{fig:varpi_clustering_statistics}). 
Galactic tides would induce a nodal precession about a warped plane that depends on the distance and eccentricity of the TNO, but they are only expected to be relevant at much larger distances ($a\gtrsim 1000$ AU, \citealt{2020CeMDA.132...12S,2021ApJ...920..148B}).
Future data from the Vera Rubin Observatory will allow us to paint a clearer picture of these trends.

\section{Conclusions}\label{sec:conclusions}
In this work, we have leveraged on a dynamically-motivated approach to give a quantitative measure of the degree of orbital diffusion of known distant trans-Neptunian objects, and thus their stability. Under the framework of Chirikov diffusion \citep{1979PhR....52..263C}, the degree of overlap of the relevant resonances with Neptune \citep{2021ApJ...920..148B} naturally divides TNOs into classes of non-diffusive, weakly-diffusive and strongly diffusive objects. Using this classification, objects whose orbits are not strongly diffusive (stable and metastable) show a clustering of both the longitude of pericenter (Fig.\ \ref{fig:varpi_clustering_statistics}) and orbital poles (Fig.\ \ref{fig:ang_mom_polar__stats-Xstable_w_EP}). Instead, the distribution of strongly-diffusive (unstable) objects is bimodal (Fig.\ \ref{fig:varpi_double-clustering_statistics__q-gtr-30_i-less-40}), and includes a similar number of both clustered and anti-clustered objects in longitude of pericenter, and a no significant warp in the mean orbital plane with respect to the invariable plane. A significant clustering in $\varpi$ around $\varpi^*\simeq 50^\circ$ emerges for semi-major axes $a\gtrapprox 150$ AU within the (meta)stable population (Fig.\ \ref{fig:varpi_clustering_statistics}, left panel), while the poles of the orbits appear warped by a few degrees even inside $150$ AU (Fig.\ \ref{fig:ang_mom_polar__binning_all_w_EP_from50AU}).

One point to keep in mind in this analysis are observational biases, which may play a role in producing an apparent alignment in the TNO population (although this has been shown to be unlikely in the overall population, \citealt{2017AJ....154...65B}, \citealt{2019AJ....157...62B}). While there is a natural bias for detecting objects with smaller perihelion distance and low-inclination, making unstable objects overrepresented in the observational sample (as they more strongly interact with Neptune), in this work we instead consider the somewhat more subtle correlation between stability and orbital clustering. In this regard, it is difficult to conjure up strong observational biases that would discriminate based on orbital stability, especially given that the orbital parameters are unknown for about a year after detection (i.e.,\ prior to extensive follow-up).
The fact that non-diffusive objects are only found to be clustered along a consistent direction, although they could in principle be discovered pointing in different directions, since unstable objects that are both clustered and anti-clustered are detected in similar proportions, suggests that there is no intrinsic observational bias that preferentially discovers clustered stable objects. In order for this correlation to arise, a survey would have to be solely directed in the direction of the cluster that we observe (approximately the intersection of ecliptic and galactic planes) while excluding all objects with low $q$, which is not the case. Instead, the stable objects come from a wide variety of surveys with very different pointing histories, and no evident coordination between biases, stability and clustering has occured \citep{2019AJ....157...62B}.
We also note that \cite{2024arXiv241018170S} recently performed a similar analysis, although they used somewhat simpler definitions of stability, and did not consider the warping of the orbital planes. They measure a $\sim3\sigma$ clustering in $\varpi$, which favors the hypothesis of a distant unseen planet \citep{2016AJ....151...22B}, although with different orbital parameters (notably, smaller semi-major axis and inclination). The degree of diffusion based on dynamical arguments performed in this work will be investigated in the context of its compatibility with the Planet Nine hypothesis in a future paper.

Another point to keep in mind is the role of small-number statistics. The fact that for very large orbital periods we only have a handful of detected objects hinders our current ability to perform robust statistical analyses.
Here, the Vera C.\ Rubin Observatory coming online in the near future is expected to reduce intrinsic observational biases in the sample and provide a significantly larger sample to analyze. This work intends to set the groundwork to do so, by utilizing the currently available dataset, which, despite its limitations, points towards orbital trends that, if confirmed, must be accounted for.

All in all, the analysis presented in this paper presents a challenge: any model of the outer Solar System that seeks out to explain its large-scale architecture must account for the measured dependence on stability of the orbital alignment of the small bodies that populate this region. Even if such observed clustering is only due to observational biases, the bias function must be able to explain why TNOs that occupy non-diffusive orbits always cluster in one direction, while diffusive TNOs are observed over a wider range of possible orbits, many of which point in another direction.
Ultimately, this paper puts forward a framework to measure this dependence of clustering on the level of diffusivity of trans-Neptunian objects, and, in the light of the upcoming Vera C.\ Rubin Observatory dataset, it appropriately places the categorization of the stability of trans-Neptunian objects into quantitative grounds.

\section{Acknowledgments}
G.P.\ wishes to thank the Barr Foundation for their financial support, and Caltech's GPS Division for their hospitality. K.\ B.\ thanks the Packard Foundation, the Caltech Center for Comparative Planetary Evolution (3CPE), and the National Science Foundation (Grant AST 2408867) for their support.

\bibliographystyle{aasjournal}
\bibliography{cas-sc-sample}

\appendix
\section{Orbital stability classification and orbital parameters}\label{apx:A}
\begin{figure}[ht!]
\centering
\includegraphics[width = 0.9\textwidth]{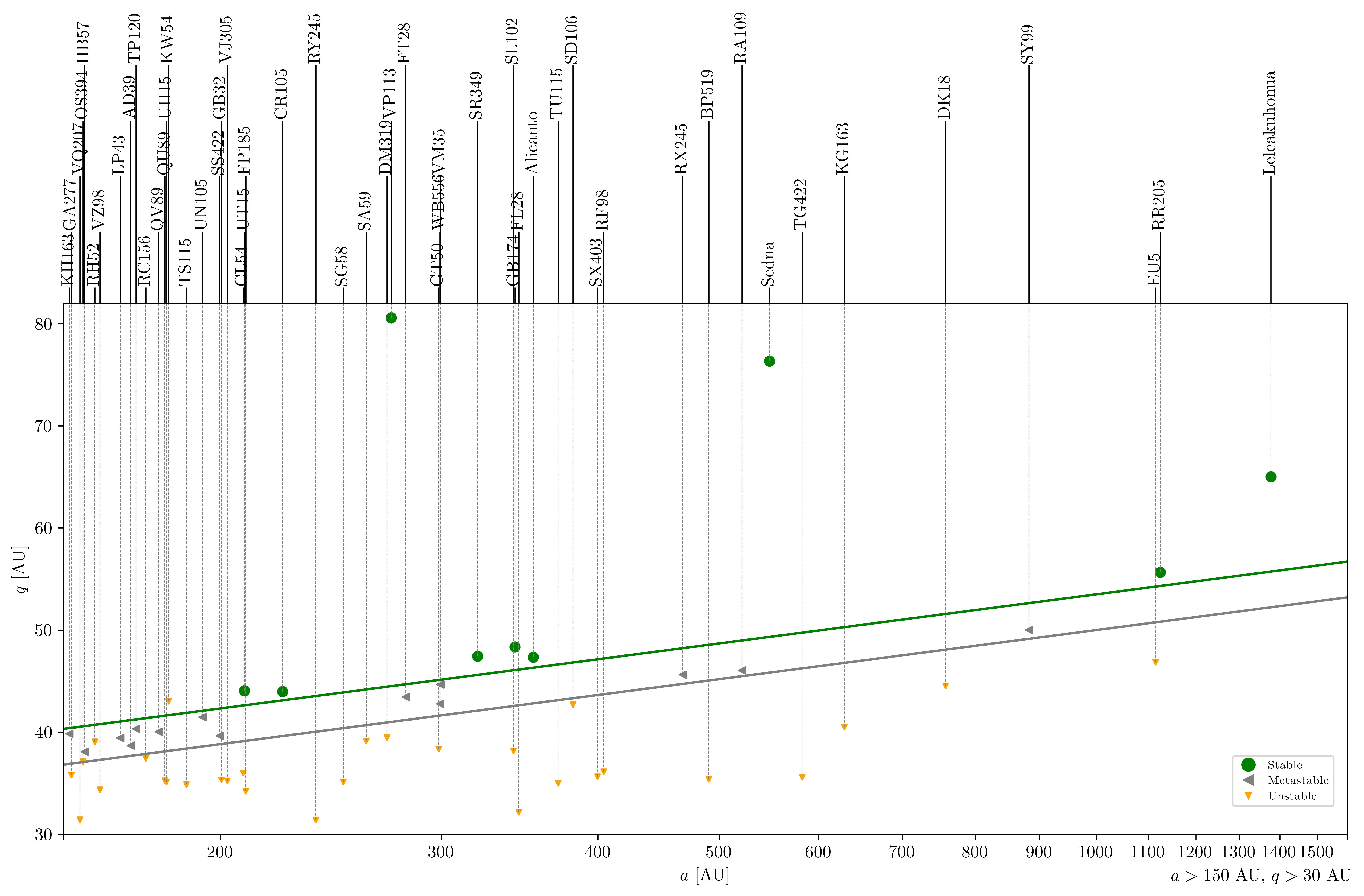}
\caption{
Perihelion distances $q$ vs.\ semi-major axis $a$ of observed TNOs with $a>150$ AU and $q > 30$ AU. TNOs are labeled by their stability as per our analysis in Sect.\ \ref{sec:Da} (see legend in the bottom-right corner). Two lines indicate a trend with $q$ which, in the region $a\gtrsim 200$ AU, separates the plane in stable, metastable and unstable bands; the values of $q$ where these divisions occur depend themselves on $a$ (see main text). For $a\lesssim 200$ AU, this distinction becomes less clear.
\label{fig:loga-vs-q__a-gtr-200_q-gtr-30__stab-through-mean-Da}}
\end{figure}

\begin{figure}[ht!]
\centering
\includegraphics[width = 0.9\textwidth]{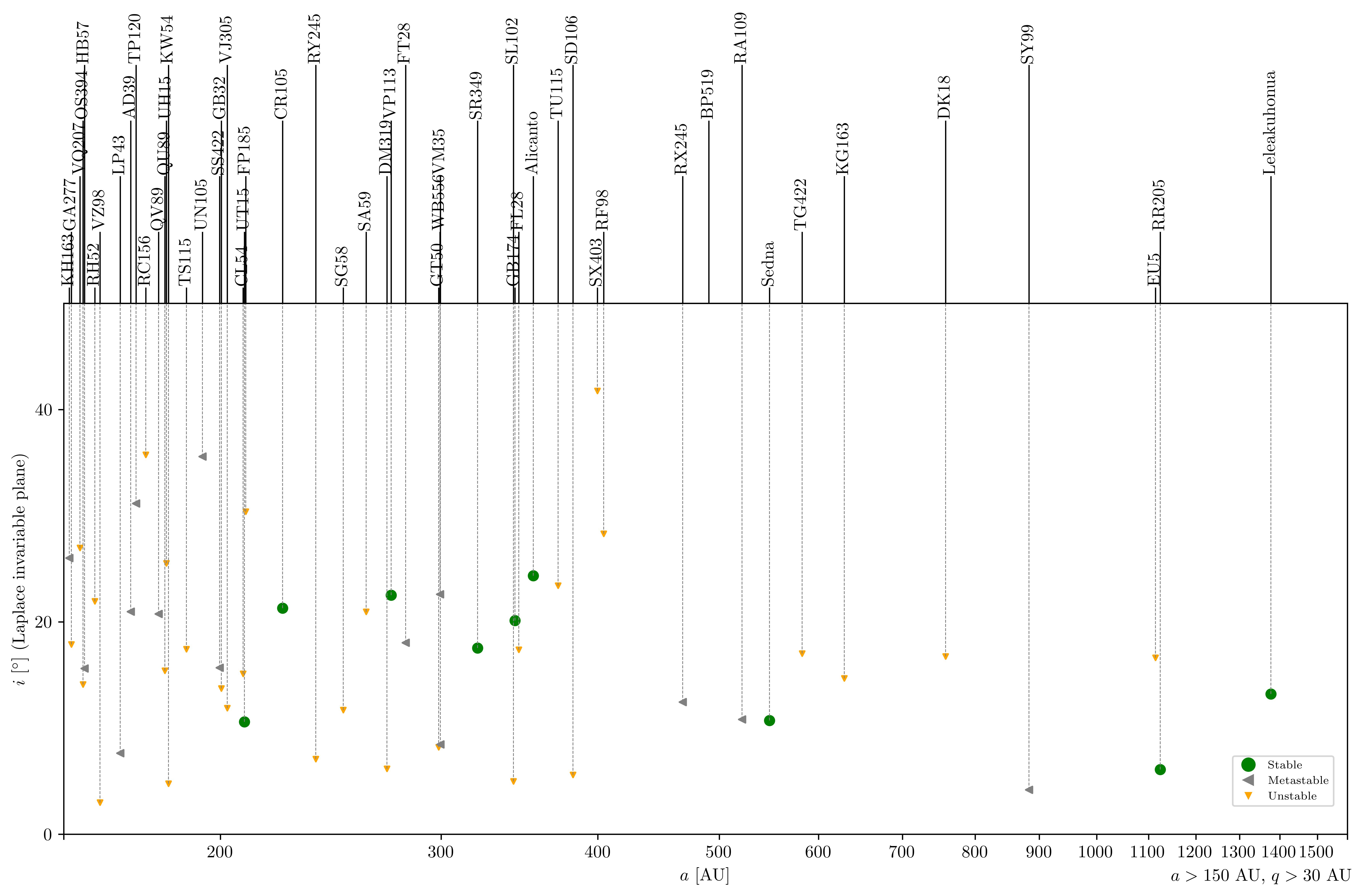}
\caption{
Similar to Fig.\ \ref{fig:loga-vs-q__a-gtr-200_q-gtr-30__stab-through-mean-Da}, showing inclination $i$ (with respect to the Laplace invariant plane) vs.\ the semi-major axis $a$ of observed TNOs with $a>150$ AU.
\label{fig:loga-vs-i_IP__a-gtr-200_q-gtr-30__stab-through-mean-Da}}
\end{figure}

\begin{figure}[ht!]
\centering
\includegraphics[width = 0.9\textwidth]{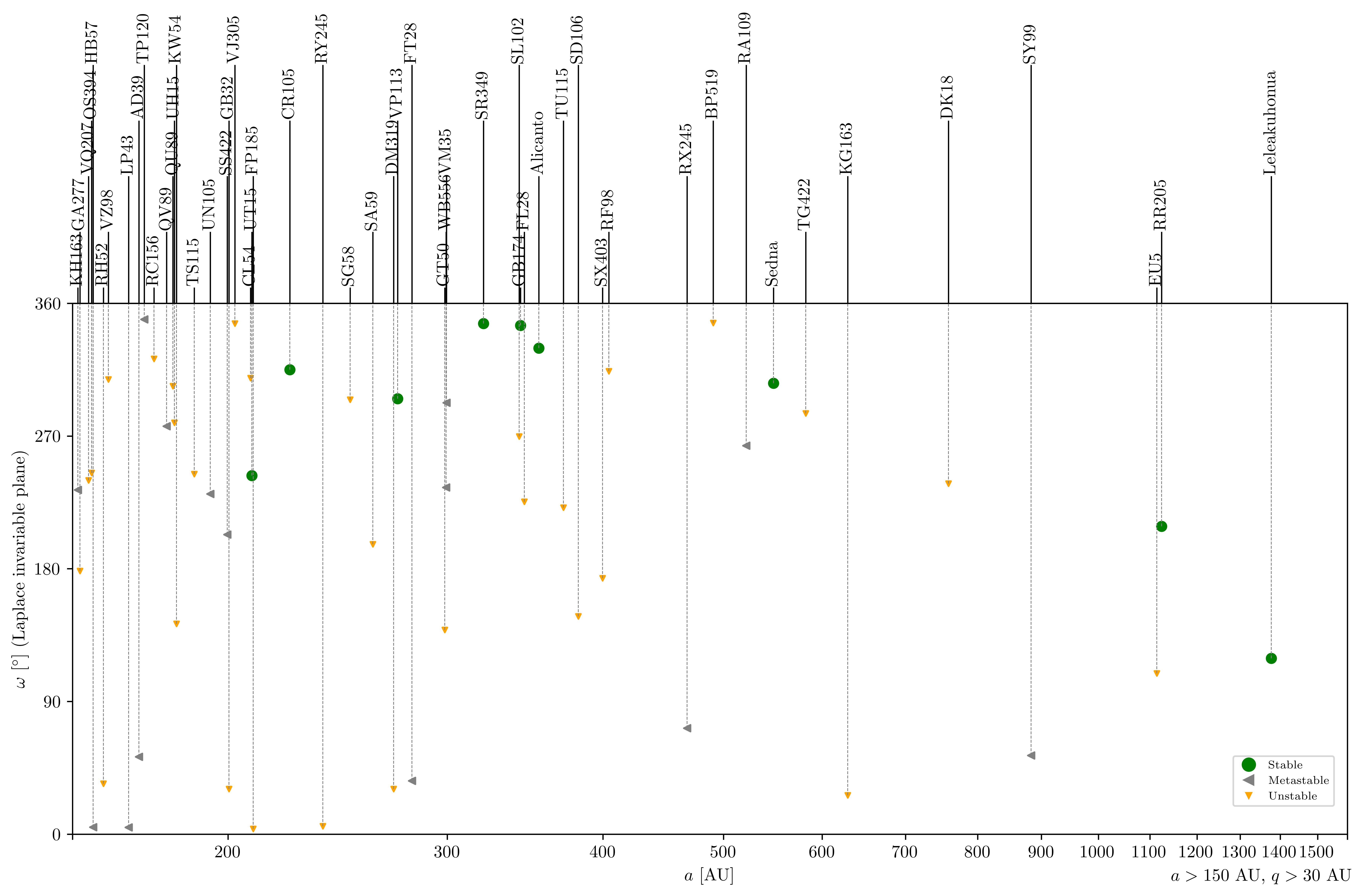}
\caption{
Similar to Fig.\ \ref{fig:loga-vs-q__a-gtr-200_q-gtr-30__stab-through-mean-Da}, showing argument of pericenter $\omega$ (with respect to the Laplace invariant plane) vs.\ the semi-major axis $a$ of observed TNOs with $a>150$ AU.
\label{fig:loga-vs-omega_IP__a-gtr-200_q-gtr-30__stab-through-mean-Da}}
\end{figure}

\begin{figure}[ht!]
\centering
\includegraphics[width = 0.9\textwidth]{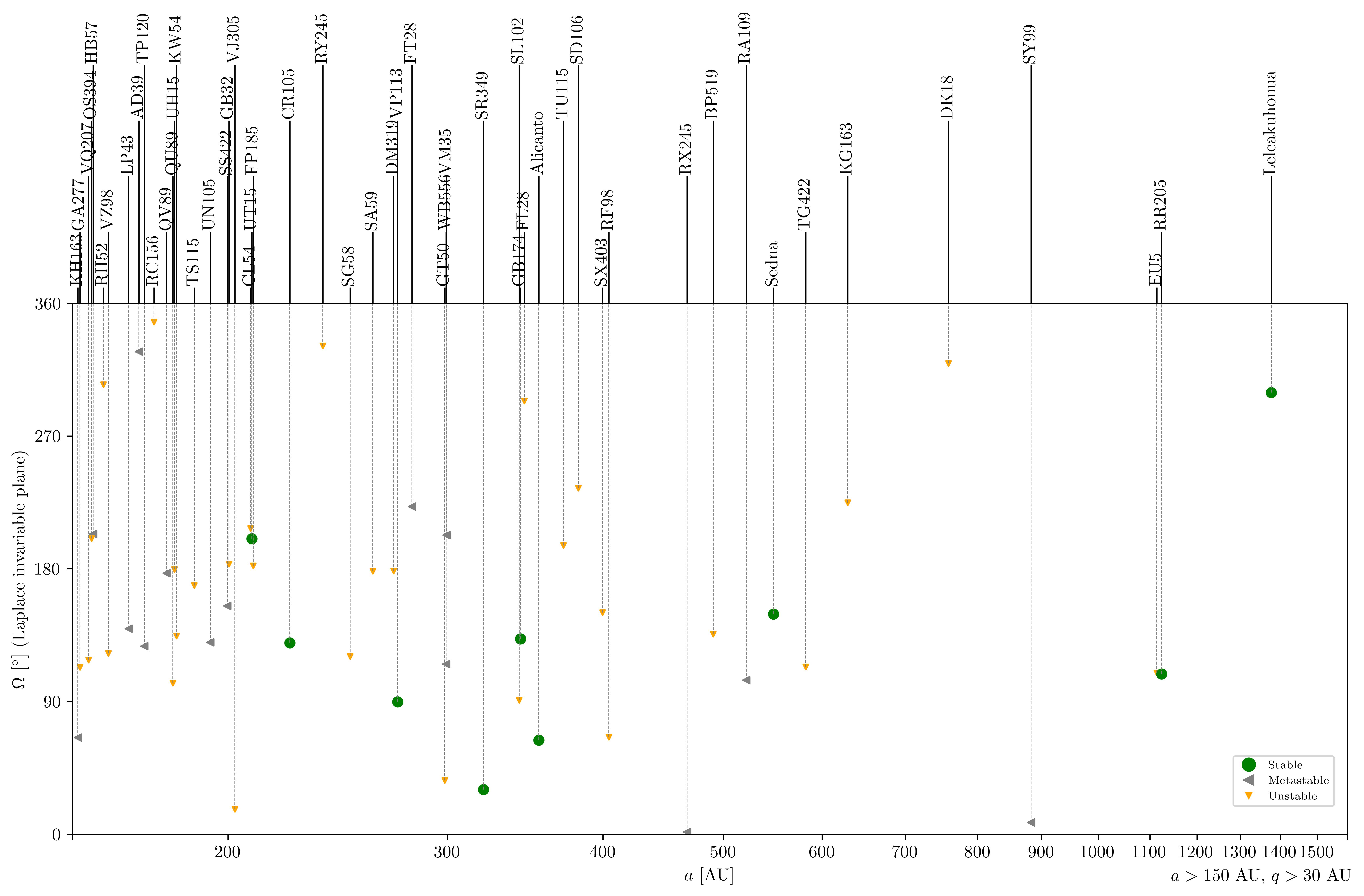}
\caption{
Similar to Fig.\ \ref{fig:loga-vs-q__a-gtr-200_q-gtr-30__stab-through-mean-Da}, showing longitude of the ascending node $\Omega$ (with respect to the Laplace invariant plane) vs.\ the semi-major axis $a$ of observed TNOs with $a>150$ AU.
\label{fig:loga-vs-Node_IP__a-gtr-200_q-gtr-30__stab-through-mean-Da}}
\end{figure}

In this Appendix, we show how the different stability properties identified in Section \ref{sec:Da} are projected onto various orbital planes. 
Figures \ref{fig:loga-vs-q__a-gtr-200_q-gtr-30__stab-through-mean-Da}, \ref{fig:loga-vs-i_IP__a-gtr-200_q-gtr-30__stab-through-mean-Da}, \ref{fig:loga-vs-omega_IP__a-gtr-200_q-gtr-30__stab-through-mean-Da}, \ref{fig:loga-vs-Node_IP__a-gtr-200_q-gtr-30__stab-through-mean-Da} show perihelion distance, inclination, argument of the pericenter and longitude of the node as a function of $\log_{10} a$, respectively.
TNOs with different orbital stability properties are found to have similar pericenter distances (Fig.\ \ref{fig:loga-vs-q__a-gtr-200_q-gtr-30__stab-through-mean-Da}), especially for $a \lesssim 200$ AU, but there may be a trend with semi-major axis for $a \gtrsim 200$ AU; see the diagonal lines separating the $q$ vs.\ $a$ plane in three bands, given by $q \simeq (16 \log_{10}(a/(1~\mathrm{AU})) + 5.5)~\mathrm{AU}$ and $q \simeq (16 \log_{10}(a/(1~\mathrm{AU})) + 2.0)~\mathrm{AU}$. A trend with $q$ is expected due to the efficiency of Neptune scattering (e.g.\ \citealt{2020CeMDA.132...12S} for a review, and \citealt{2021ApJ...920..148B,2024MNRAS.527.3054H}), but it starts to break down for $a\lesssim 200$ AU. As an example, KW54 is above the green ``stability line'' but is rather diffusive; note also that its inclination is relatively low, so the co-planarity approximation from \cite{2021ApJ...920..148B} is valid. However, the results from \cite{2021ApJ...920..148B} are applicable at large semi-major axes, and do not apply in the $a \lesssim 200$ AU regime. As already evident in Figure \ref{fig:ang_mom_polar__a-gtr-200_obs_with-EP}, Figures \ref{fig:loga-vs-i_IP__a-gtr-200_q-gtr-30__stab-through-mean-Da} and \ref{fig:loga-vs-Node_IP__a-gtr-200_q-gtr-30__stab-through-mean-Da} show that objects with various stability properties are found to coexist in the $i$ and $\Omega$ planes, as is the case for $\omega$ (Fig.\ \ref{fig:loga-vs-omega_IP__a-gtr-200_q-gtr-30__stab-through-mean-Da}).

\end{document}